\documentclass[sigconf,9pt]{acmart}
\usepackage[ruled,vlined]{algorithm2e}
\usepackage{caption}
\usepackage{subcaption}
\usepackage{balance}
\usepackage{xspace}
\usepackage{graphicx}

\AtBeginDocument{%
  \providecommand\BibTeX{{%
    \normalfont B\kern-0.5em{\scshape i\kern-0.25em b}\kern-0.8em\TeX}}}

\setcopyright{acmcopyright}
\copyrightyear{2023}
\acmYear{2023}
\acmDOI{10.1145/3625687.3625812}

\renewcommand{\paragraph}[1]{\vspace{2pt plus 0pt minus 2pt}\noindent{\bfseries #1}}

\newcommand{\systemname}{HandFi}
\newcommand{\systemnames}{HandFi's}
\newcommand{\networkname}{HandNet}

\begin{document}

\title{Construct 3D Hand Skeleton with Commercial WiFi}

\acmYear{2023}\copyrightyear{2023}
\setcopyright{acmlicensed}
\acmConference[SenSys '23]{ACM Conference on Embedded Networked Sensor Systems}{November 12--17, 2023}{Istanbul, Turkiye}
\acmBooktitle{ACM Conference on Embedded Networked Sensor Systems (SenSys '23), November 12--17, 2023, Istanbul, Turkiye}
\acmPrice{15.00}
\acmDOI{10.1145/3625687.3625812}
\acmISBN{979-8-4007-0414-7/23/11}

\author[Sijie Ji, Xuanye Zhang, Yuanqing Zheng, Mo Li]{Sijie Ji$^{\ast}$, Xuanye Zhang$^{\ast}$, Yuanqing Zheng$^{+}$, Mo Li$^{\dagger\ast}$ }
\affiliation{%
  \institution{ $^\dagger$The Hong Kong University of Science and Technology, \\$^\ast$Nanyang Technological University, $^+$The Hong Kong Polytechnic University
  }
  \country{Email:\{sijie001, c200212\}@ntu.edu.sg, yqzheng@polyu.edu.hk, lim@cse.ust.hk}
}

\def \authors{Sijie Ji, Xuanye Zhang, Yuanqing Zheng, Mo Li}


\begin{abstract}
This paper presents \systemname, which constructs hand skeletons with practical WiFi devices. Unlike previous WiFi hand sensing systems that primarily employ predefined gestures for pattern matching, by constructing the hand skeleton, \systemname{} can enable a variety of downstream WiFi-based hand sensing applications in gaming, healthcare, and smart homes. 
Deriving the skeleton from WiFi signals is challenging, especially because the palm is a dominant reflector compared with fingers. 
\systemname{} develops a novel multi-task learning neural network with a series of customized loss functions to capture the low-level hand information from WiFi signals. During offline training, \systemname{} takes raw WiFi signals as input and uses the leap motion to provide supervision. During online use, only with commercial WiFi, \systemname{} is capable of producing 2D hand masks as well as 3D hand poses.
We demonstrate that HandFi can serve as a foundation model to enable developers to build various applications such as finger tracking and sign language recognition, and outperform existing WiFi-based solutions. Artifacts can be found: \hyperlink{https://github.com/SIJIEJI/HandFi}{https://github.com/SIJIEJI/HandFi}
\end{abstract}







\maketitle

\section{Introduction}

Hand sensing plays a crucial role in human-computer interaction, enabling a  broad range of applications in video games, education, and healthcare. These applications further create new opportunities for people with communication disabilities to interact with devices and others.
Numerous hand-sensing systems have been developed, wherein wearable-based solutions offer high accuracy but are limited by their cumbersome nature \cite{xu2021limu}. On the other hand, contact-free solutions offer greater flexibility and versatility. However, acoustic-based solutions are susceptible to environmental noise and have a limited sensing range \cite{wang2020push,yang2023voshield}, while computer vision-based solutions may raise privacy concerns. In contrast, RF-based solutions offer a wider sensing range, preserve privacy and are robust to illumination variation.  Among these technologies, WiFi-based solutions are promising to repurpose ubiquitous low-cost WiFi devices for contact-free hand sensing.

\begin{figure}[t!]
  \centering
  \includegraphics[width=\linewidth]{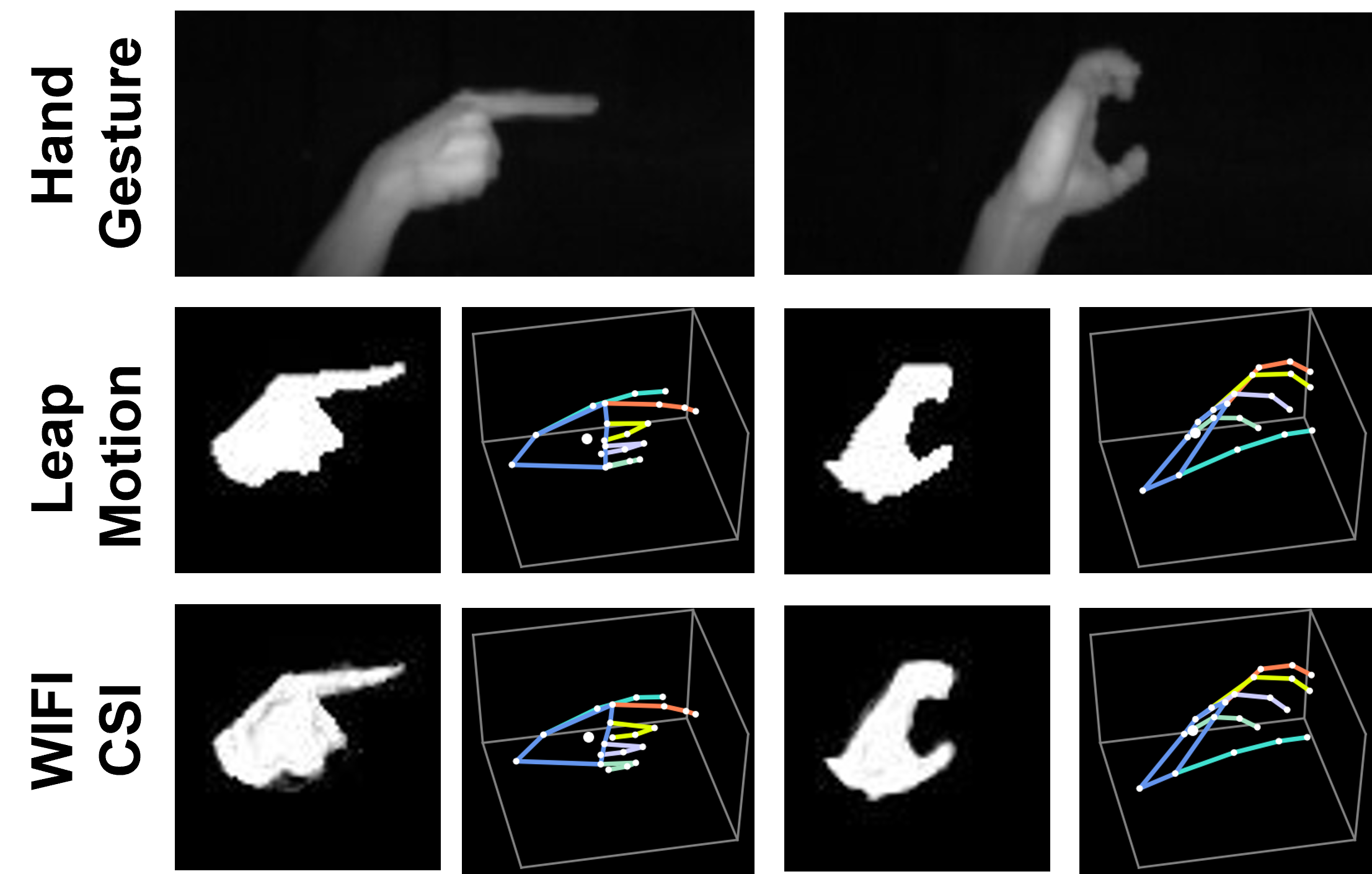}
  \caption{Leap Motion Output VS \systemname{} Output.}
\vspace{-0.4cm}
  \label{f:keyfigure}
\end{figure}

However, due to the low spatial resolution of WiFi signals and the small scale of a hand, directly modeling the radio reflection from a hand is challenging. Therefore, existing WiFi-based solutions rely on WiFi signal patterns caused by hand motions to their corresponding hand gestures~\cite{shang2017robust,zhang2020wisign,ma2018signfi,xing2022wifine,li2016wifinger,tan2016wifinger}, or employ geometric constraints to track variations in signal propagation for hand tracking~\cite{sun2015widraw,tan2020enabling,zhang2018letfi,yu2018qgesture,wu2020fingerdraw}. 
None of the existing WiFi-based hand-sensing systems directly models the relationship between the hand skeleton and the reflected signals of interests. As a result, specific patterns or models need to be identified and built for different downstream  applications. Thus, these gesture recognition systems are typically limited to a small pre-defined set of hand gestures, while signal propagation modeling-based hand tracking solutions are vulnerable to noise and environmental changes.  Nonetheless, the existing WiFi-based hand sensing systems show that WiFi signals do carry information about human hands and hand movements. In addition, WiFi has the potential to achieve millimeter-level sensing resolution~\cite{wang2016human,jiang2020towards}. 
The challenge lies in how to separate the signal of interest of the target reflector from the environment-related multi-path and hardware imperfection-induced noise, which are often nonlinearly superimposed.
On the other hand, deep learning is particularly well-suited for extracting the signal of interest from the nonlinear superimposed high-dimensional data.  A deep learning model could potentially  learn the intricate relationships of the superimposed WiFi CSI and extract the reflected signal from the hand through proper design of neural networks.
The question remains whether it is possible to extract rich hand semantic information from WiFi signals and achieve vision-like results such as 3D hand pose construction with the novel design of neural networks. If such a result can be achievable, we will be able to build various downstream applications directly and integrally without being limited by pre-defined hand gestures.

This paper explores the possibility of obtaining vision-like results solely using commercial WiFi by presenting \systemname{}. \systemname{} is capable of constructing the shape and skeleton of a hand simultaneously. Figure~\ref{f:keyfigure} presents the output of \systemname{} compared with the output of a commercial depth camera (leap motion). The hand shape is represented by a binary matrix-based hand mask, while the hand skeleton is represented by a set of vectors indicating the 3D coordinates of 21 key joints of a hand. 
The core of \systemname{} is a novel deep neural network called \networkname{}, which is trained in an end-to-end manner with cross-modality supervision using labels obtained from the depth camera. Once a model is trained, \systemname{} can directly infer hand shape and skeleton, which can be used to support a variety of downstream applications with flexible extension. For example, finger tracking can be directly enabled by tracking the coordinate of an index finger in 3D hand skeleton. Based on the 3D hand skeleton, a new hand gesture can be defined and added in a more efficient rule-based manner (e.g., relative position of key joints) rather than a data-driven manner which would otherwise require cumbersome new data collection and model retraining. 


Developing such a hand-sensing foundation model capable of constructing vision-like shape and skeleton information solely using commercial WiFi entails two major technical challenges, even with the assistance of a depth camera during model training. First, the palm is a dominant reflector compared to the fingers. It is challenging to separate different scales of reflectors from the received signal and model the relationship between the dominant reflector (the palm) and other reflectors (the fingers) to further understand the structure of the hand.
Second, the hand's varying positions and changes in the ambient environment can lead to distinct CSI multi-path profiles. As such, the developed model must be capable of accurately extracting the signals of interest from the hand while remaining robust to any other changes.

To address the aforementioned challenges, we develop \networkname{}, a novel symbolically-constrained multi-task learning framework. HandNet first adds one more modality (hand mask) to learn together with the hand pose by sharing the same encoder. 
This additional modality provides dense supervision, constraining the learning process and balancing the information between the dominant reflector (palm) and other reflectors (fingers).
Simultaneously, the encoder is designed to preserve the phase information (distance information) of the CSI plus multi-scale feature extractors to exploit reflectors of different sizes that are of interest. Further, a set of symbolically constrained loss functions based on the defined parameterized hand model are adopted to ensure the reconstruction of anatomically plausible skeletons. In addition, a domain generalization method is introduced during HandNet training and we collect a comprehensive training data set to train the model. The model is encouraged to learn only the features representing the hands of interest that are common across different hand positions and environments. 
These techniques put together allow us to construct 3D hand skeleton with practical WiFi devices.
\systemname{} is evaluated comprehensively in various usage scenarios and we summarize the results and contributions as follows :
\begin{itemize}
    \item We develop \systemname{} which is capable of constructing 2D hand mask and 3D hand pose with commercial WiFi devices. 
    \systemname{} achieves a 91\% overlap with ground truth for 2D hand masks and a 2.07 cm joint error for 3D hand skeletons, sufficiently accurate for various HCI applications.
    \item \systemname{} can serve as a foundation model for the development of a variety of downstream hand-sensing applications. For example, finger-level sensing applications developed based on \systemname{} outperform existing WiFi-based sensing systems.
    \item \systemname{} has been prototyped and evaluated under various conditions, including different sensing ranges, hand positions, different users, and occlusion scenarios. \systemname{} can work under occlusion and offers an extended field-of-view and sensing range.
\end{itemize}

\section{Overview}

\begin{figure}[t]
  		\centering
		\includegraphics[width=0.85\linewidth]{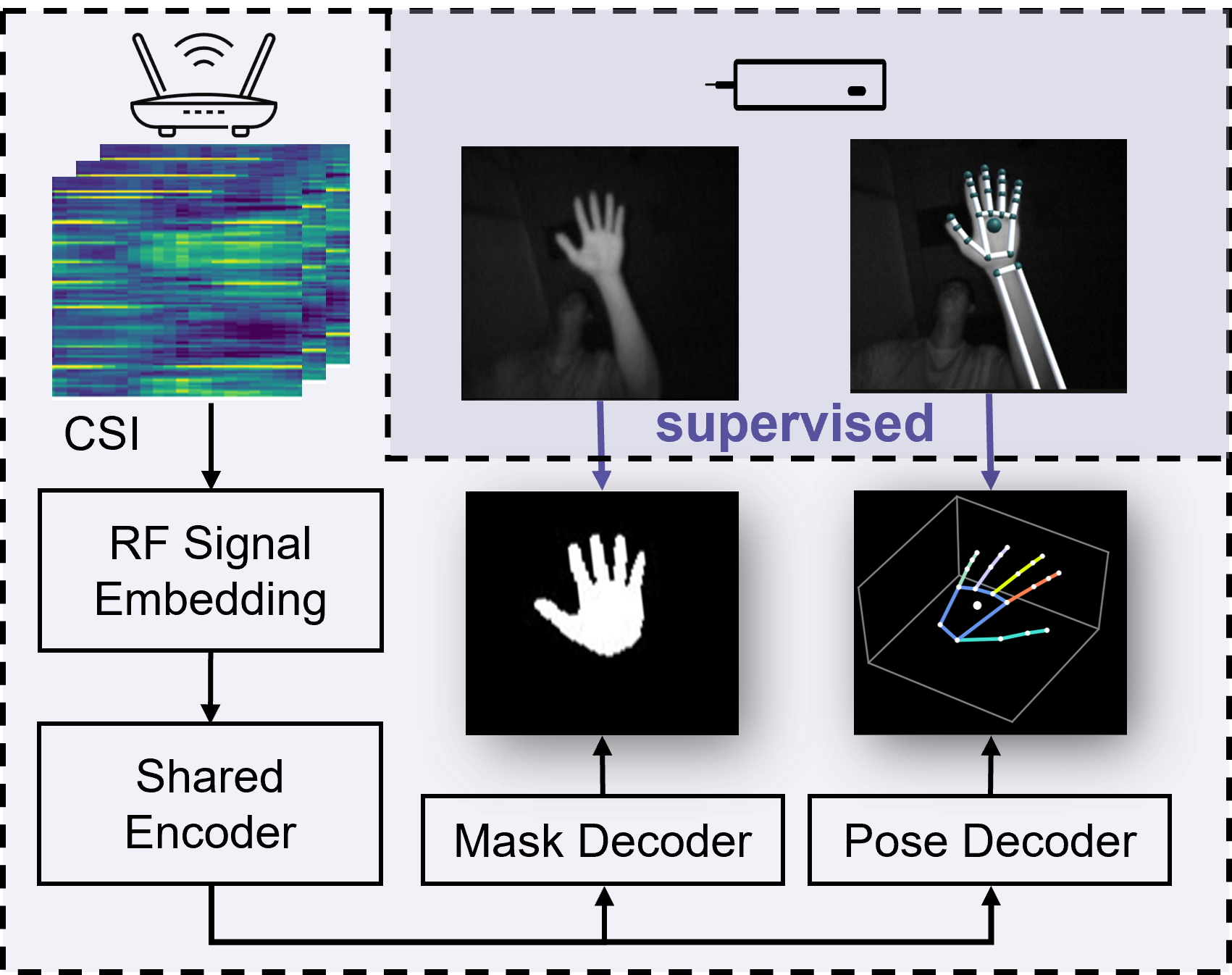}
		\caption{System overview.}
		\label{f:system}
\end{figure}

\systemname{} takes as input the channel state information (CSI) that measures the environment by multiple antennas and then goes through \networkname{} 
to construct 2D hand mask and 3D hand pose as illustrated in Figure~\ref{f:system}. The depth camera is only used in the training stage to provide ground truth labels.
The core of \systemname{} is the design of a multi-task learning network, \networkname{} (Figure~\ref{f:handnet}), including four main components. The first component is the RF signal embedding layer to deal with complex-valued CSI from different antenna streams, aiming to preserve both amplitude and phase information of CSI. 
The second component consists of a shared deep multi-scale encoder, which is designed to squeeze and exploit the signals of interest from the palm and finger reflectors contained within the high-dimensional noisy CSI and then transform relevant deep hand semantic features into a latent space. Two task-specific decoders then reconstruct the 2D hand mask and 3D hand pose from the latent deep hand semantic features. This decoding process provides complementary information if a WiFi frame is unable to encode all parts of the hand.  
To make the system deployable in unseen hand positions and environments, HandFi further adopts a domain generalization technology. In the end, a wide range of applications can be built on top of the inferred hand mask and hand pose. For example, one can encode different gestures as a one-hot vector and add one fully connected layer as a classification head to classify the gestures in a flexible way. New gestures then can be added by updating the one-hot vector. Likewise, one can focus on one of the fingers and enable finger tracking.

\section{Methodology}

This section outlines the methodology of \systemname{} and its core multi-task learning-based framework, \networkname{}.
Although the range resolution of WiFi is low (with a range resolution of only 15 meters for a 20MHz bandwidth), the measurement granularity of CSI is high and sufficient to distinguish subtle multipath profile variation over a short distance. 

For exmaple, the CSI data obtained at 5GHz with a wavelength $\lambda=5.7cm$, and $\pi/3$ phase measurement granularity is capability of distinguishing 5mm distance resolution~\cite{wang2016human}. However, the imperfections in WiFi hardware can induce mixture phase rotations to the CSI, making it difficult to separate the phase rotation caused by the target reflector. Therefore, we resort to the deep learning technology, which is good at extracting features from a high-dimensional data to model the relationship between hand pose and CSI data. 
To establish a one-to-one relationship with ground truth label and incorporate symbolic prior knowledge, we first introduce the proposed hand model. Next, we explicitly embed the complex-valued CSI through point-wise group convolution, which prevents information loss and preserves the physical information of CSI. The embedded CSI is then input into a multi-scale perception encoder that extracts features from different domains with different scales and transforms them into a deep hand semantic space. This allows us to focus on signals of interest and filter out irrelevant noise. Two task-specific decoders are connected to the same encoder, which works in parallel to reconstruct the 2D hand mask by 0-1 classification and 3D hand pose by regression, respectively. A set of customized loss functions is utilized for both tasks. The back-propagation of the 2D hand mask reconstruction task regularizes the learned latent space of the shared encoder, which then benefits the 3D hand pose task. Further, the domain generalization technique is applied during training to learn position-agnostic latent features for practical concerns.

\begin{figure}[t]
  \centering
  \includegraphics[width=0.5\linewidth]{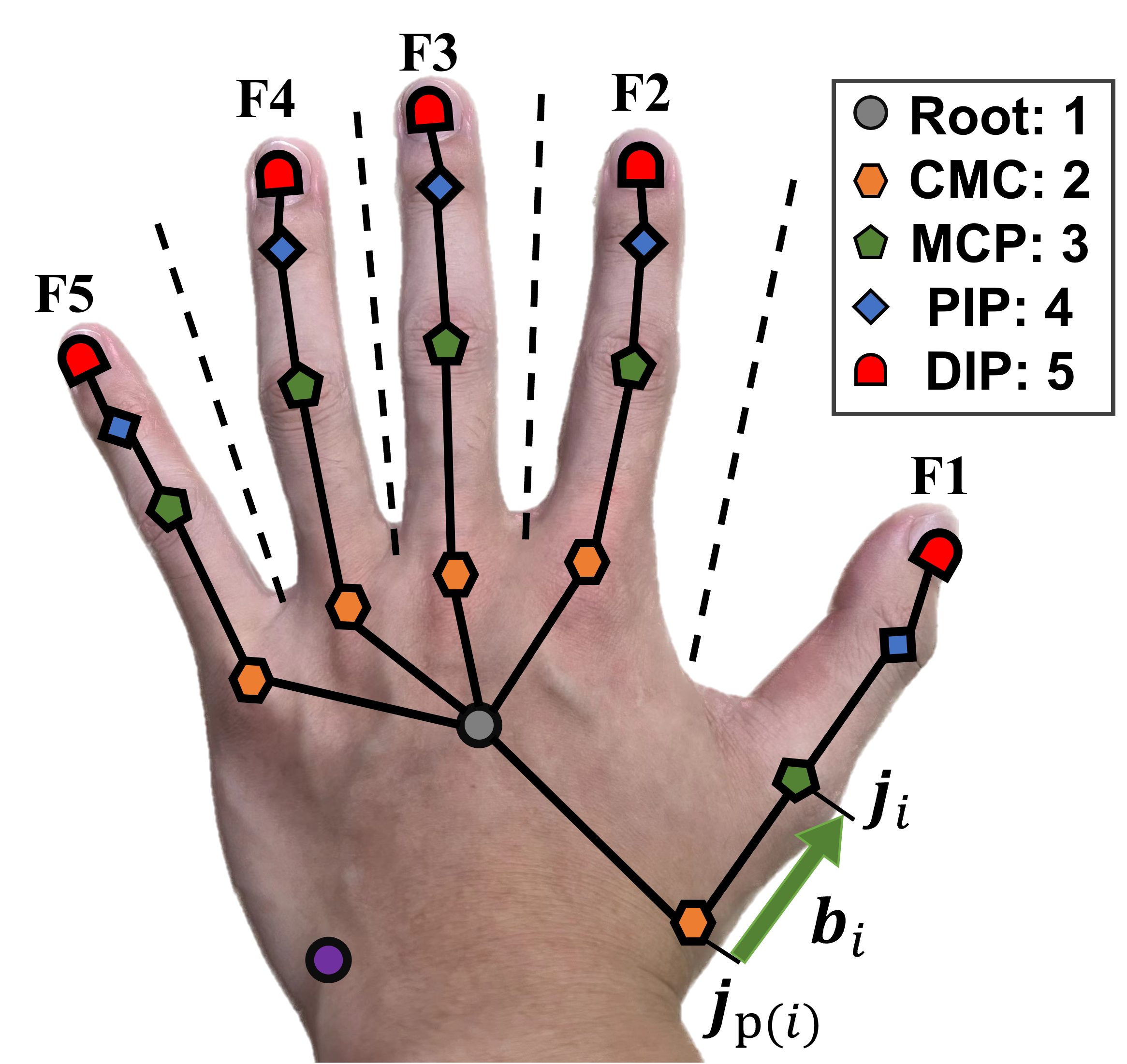}
  \caption{The hand model.}
  \label{f:handmodel}
\end{figure}
\subsection{Hand Modeling}\label{s:hand model}
In order to establish one-to-one correspondences with ground truth labels for supervised learning, \systemname{} first symbolically models the hand. The depth camera returns 24 key joints of the hand, with 20 joints representing the fingers, one joint indicating the center of the palm and three joints denoting the wrist (as shown in the top right corner of Figure~\ref{f:system}). The depth camera provides sufficiently accurate results that we take as ground truth. We keep the 20 finger joints along with the palm joint, while discarding the remaining joints to build our hand model as shown in Figure~\ref{f:handmodel}. Our model differs from the well-recognized 21-joint hand model in the computer vision (CV) domain~\cite{qian2014realtime}, where the wrist point is used as the root point. Instead, we choose the palm point as the root point. This is not only because the palm is a dominant reflector, but also because the depth camera uses it as the base to estimate fingers and the wrist, providing a more accurate ground truth for locating the center of the palm.
The 3D hand pose is defined by a set of coordinates $\mathbf{j_{i}} = \left(x_{i}, y_{i}, z_{i}\right)$, which describe the locations of $J$ keypoints in 3D space, i.e., $i \in \left[ 1,...,J\right]$ where $ J = 21$ in our case. The joint matrix $\textbf{\textit{J}} \in \mathbb{R}^{21\times3}$. 
The depth camera does not directly provide the hand mask. However, \systemname{} extracts the hand mask from the raw image frame captured by the depth camera. The hand mask is represented as a 0-1 matrix $\textbf{\textit{M}} \in \mathbb{R}^{114\times114}$, where the areas with and without the hand are annotated respectively. We model the 2D hand mask and 3D hand pose to direct \networkname{} to learn the gesture-independent hand shape and hand skeleton. 


%

Since the 21 joints follow the hand's anatomy and the hand is a complex articulated object, one can further formulate them to facilitate the incorporation of prior knowledge in subsequent learning procedures. Specifically, the order of the joints is placed elaborately. The hand joints are grouped by fingers. Each finger consists of sequential joints that provide constrained motion, referred to as a kinematic chain - Carpometacarpal Joint (CMC) , Metacarpophalangeal Joint (MCP), Proximal Interphalangeal Joint (PIP) and Distal Interphalangeal Joint(DIP), denoted as the respective set $\textbf{\textit{F}}_{q}$, $q \in \left[ 1,...,5\right]$ and $\textbf{\textit{F}} \in \mathbb{R}^{4\times3}$. 
Except for the root joint $\mathbf{j_{1}}$, each joint has a parent $p(i)$ and we define a bone as a vector pointing from the parent joint to its child joint, $\mathbf{b_{i}} = \mathbf{j}_{i+1} - \mathbf{j}_{p(i+1)}$, so $\left [ \mathbf{b_{1}},...,\mathbf{b_{20}} \right ] = \textbf{\textit{B}} \in \mathbb{R}^{20\times3}$. The bones are named according to the child joint. For example, the bone connecting CMC to MCP is called MCP bone. Intuitively, the CMC bones share one root joint $\mathbf{j_{1}}$ are those that lie within and define the palm. We define the CMC bones $\mathbf{b_{1}},...,\mathbf{b_{5}}$ to correspond to the fingers $\textbf{\textit{F}}_{1},...,\textbf{\textit{F}}_{5}$. Figure~\ref{f:handmodel} visualizes the hand model (The purple joint is excluded from the model and learning process, and is solely utilized for plotting the 3D hand pose for illustration purposes) . 
\vspace{-0.2cm}

\subsection{RF Signal Embedding}
Specular reflections in RF signals make it difficult to determine if each frame contains all the necessary components of the hand. We aim to retain as much information as possible during signal processing, trusting in the power of deep neural networks to filter out irrelevant features. To achieve this goal, we design an RF signal embedding layer that preserves the physical information carried by CSI, serving as an adapter between complex-valued signals and real-valued deep learning building blocks. Notably, it is important to preserve the phase information, particularly as it represents the corresponding distance of hand reflectors.
Unlike images where all pixels have the same magnitude (0-255), CSI values are much more dynamic. For example, the pathloss grows exponentially with distance, and the CSI, $h=a+bi$, is complex-valued, containing both magnitude and phase information of the channel coefficient. Traditional normalization methods, such as scaling the values to a certain range [0,1], are not appropriate for CSI. Therefore, we first normalize CSI by the average power of each packet:
\begin{equation}
    \hat{\textbf{\textit{h}}} = \textbf{\textit{h}} / ( \sum_{i=1}^{F} \left \| h_i \right \| /F) 
\end{equation}
where \textbf{\textit{h}} is a vector of CSI of one packet, the length of the vector depends on the number of subcarrier $F$. 
Then, we separate the real and imaginary parts of the normalized CSI to form a real CSI matrix and an imaginary CSI matrix accordingly. We then stack them together to form a tensor denoted as $\textbf{\textit{H}} \in \mathbb{R}^{F\times T \times 2}$,  where $F$ is the number of subcarriers and $T$ is the number of packets. After pre-processing, CSI data meets the requirements of the operators of deep learning.

\begin{figure*}[t!]
  \centering
  \includegraphics[width=0.98\linewidth]{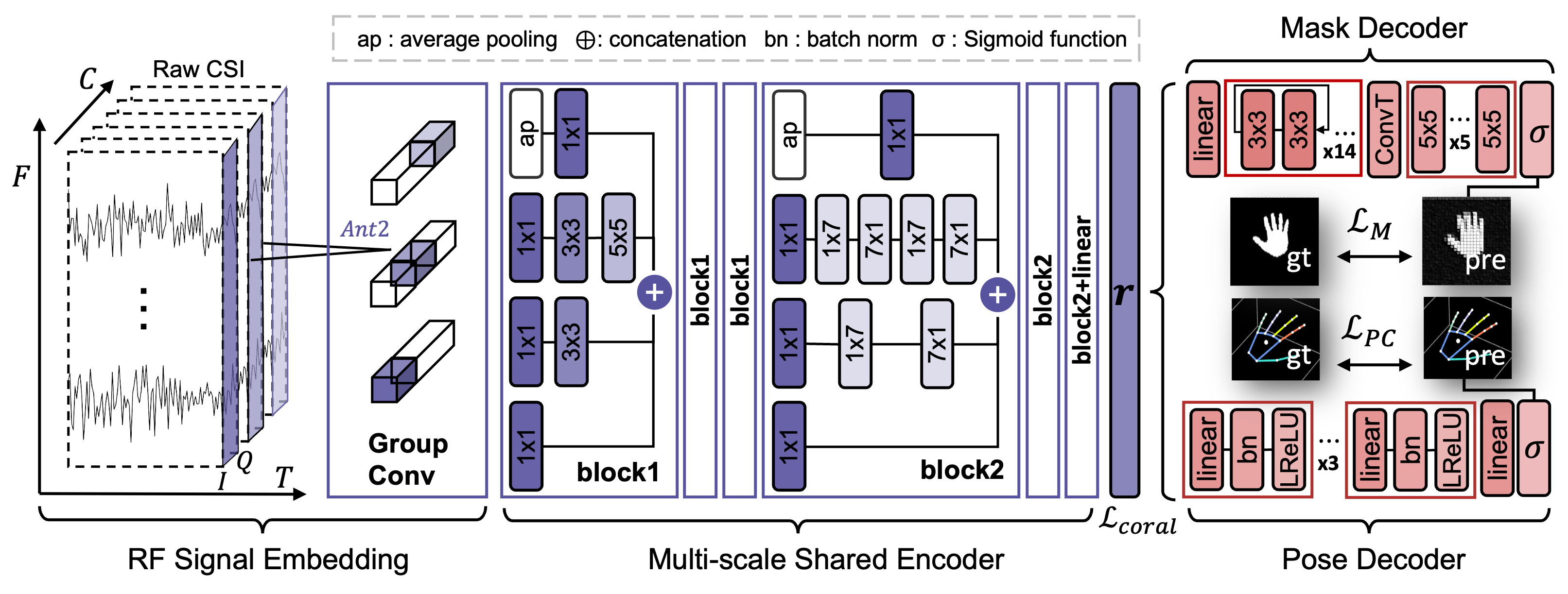}
  \caption{Architecture of HandNet.}
  \label{f:handnet}
\end{figure*}

Although we separate the real and imaginary parts to adapt to existing deep learning frameworks, we do not want to lose the physical information contained in the complex-valued CSI. For example, the absolute value of $a$ and $b$ is the amplitude of the CSI and the ratio between $a$ and $b$ is the phase of the CSI. 
Inspired by the CLNet~\cite{ji2021clnet}, we 
explicitly embed the CSI by making the first layer of \networkname{} as 1 $\times$ 1 point-wise convolution without bias term such that:
\begin{equation}
E_{i} = w_{i}a + w_{i}b
\end{equation}
where $w_{i}$ is the weight of the convolution filter and $i$ is the number of filters. Such embedding preserve the phase information by maintaining the ratio between $a$ and $b$, and the amplitude information is simply scaled by the $w_{i}$ parameter. Unlike CLNet which only considers one CSI matrix, a wireless sensing system typically contains multiple antennas that render multiple CSI matrices. Hence, we stack different CSI matrices across the channel dimension,  $\textbf{\textit{H}} \in \mathbb{R}^{F\times T \times (2 \times Ant)}$, where $Ant$ denotes the number of spatial streams, and apply group convolution that embeds each pair of CSI matrices from the same antenna with learnable weights to preserve the physical information of CSI as the CLNet does. Figure~\ref{f:handnet} depicts the RF signal embedding operation.

\subsection{Multi-task Learning} 
The RF signals are now organized into a time-frequency-spatial tensor with the physical information preserved to feed in the shared encoder with two task-specific decoders as shown in Figure~\ref{f:handnet}.
Multi-task learning improves the performance of tasks by utilizing the limited training samples to learn generic features that benefit from the effect of regularization brought by parameter sharing, which has been successfully adopted in many domains~\cite{yang2016deep}, including hand pose estimation in CV domain~\cite{zhang2021hand}. 
In \systemname{}, the ground truth 2D hand mask and 3D hand pose are acquired from the same egocentric camera. Although they may have different distributions and unequal information, they are inherently paired and possess intrinsic correspondence, enabling them to be learned jointly. By harnessing the dense supervision provided by the 2D hand mask, the information regarding the overall hand structure can be propagated to the task of reconstructing the 3D hand pose. This process aids in disambiguating the hand pose amidst the complexity of the WiFi signals.








\subsubsection{Shared Multi-scale Perception Encoder}
The information pertaining to the hand is intricately encoded within high-dimensional and noisy WiFi signals, where the palm assumes the role of the dominant reflector, while the fingers exhibit comparatively smaller reflections. To capture deep semantic information of the hand across diverse scales, \networkname{} has devised an encoder equipped with varying receptive fields,  allowing \networkname{} to extract and integrate information from different feature dimensions. The job of encoder is to exploit and extract deep semantic information of the hand regardless of noise and other irrelevant CSI patterns. 
To address this problem, we equip the network with modified InceptionV3~\cite{szegedy2016rethinking} blocks to enhance feature extraction across multiple scales.
Specifically, \networkname{} creates four learning pathways (from top to bottom of block 1 and block 2), as shown in Figure~\ref{f:handnet}. After RF signal embedding layer, $\textbf{\textit{H}} \in \mathbb{R}^{F\times T \times Ant}$, where $Ant$ represents the number of antenna streams ($Ant = 3$ in our case). 
The tensor's height, width, and depth respectively represent information in the frequency sub-channels, time domain, and spatial domain.
The four pathways aim to extract features of different dimensions and scales. In particular, the combination of average pooling (AP) and 1x1 convolution filters are specifically designed to capture spatial features by condensing all time-frequency features into a single numerical value through AP and then filtering responses throughout the depth of the tensor. The second pathway is intended to extract fused global time-frequency information using a combination of filters of different scales in a large receptive field. The 1x7 filter focuses solely on the time domain, the 7x1 filter focuses solely on the frequency domain, and the 1x1 filter is used to increase the tensor dimension for improved learning. The third pathway is designed to focus on local time-frequency features that are complementary to those extracted by the second pathway. Similarly, the fourth pathway is designed to focus on local spatial features that complement those extracted by the first pathway. All four pathways are concatenated to fuse features with different scales of the time-frequency-spatial domains. At the end, the multi-scale encoder will obtain a deep representation of $\textbf{\textit{H}}$, CSI, denoted as $\textbf{r}$.

\subsubsection{2D Hand Mask Generation}

A task-specific decoder is devised to generate the 2D mask $\mathbf{\hat{M}}$ from the deep hand representation $\textbf{r}$ using ground truth 2D hand mask supervision, as illustrated in the top right of Figure~\ref{f:handnet}. The decoder is composed of 14 residual blocks~\cite{he2016deep}. These residual blocks are designed to be more sensitive to gradient changes and to prevent feature loss in the mask decoder. A transport convolution layer with five 5x5 convolution blocks is then used to rescale the feature map to a fixed hand mask size for the supervision training. Instead of directly regressing a hand mask from CSI, we consider the hand mask generation as a pixel-level classification problem and thus adopt binary cross entropy (BCE) loss rather than mean square error (MSE) loss such that:
\begin{equation}
\mathcal{L}_{BCE}=-\frac{1}{n} \sum_{i=1}^n[\mathbf{m} \log (\widehat{\mathbf{m} })+(1-\mathbf{m}) \log (1-\widehat{\mathbf{m}})]
\end{equation}
where $n$ is the number of training samples of the hand mask. Compared to small value gradient of MSE loss, the BCE loss return gradient is proportional to the difference between the prediction and the truth and improve the network ability to distinguish the hand class. 
Inspired by Focal Loss~\cite{lin2017focal}, we add a factor $(1 - \mathbf{m}_{t})^{\gamma}$ to $\mathcal{L}_{BCE}$ for the purpose of focusing on hard, misclassified elements, which in our case is the boundary part of the hand and the fingers areas. For notational convenience, we define $\mathbf{m}_{t}$:
\begin{equation}
\mathbf{m}_{\mathrm{t}}= 
\begin{cases}
\widehat{\textbf{m}} & \text { if  \textbf{m}=1 }\\ 1-\widehat{\textbf{m}} & \text { otherwise }
\end{cases}
\end{equation}
So, our loss for 2D hand mask generation is defined as:
\begin{equation}
\mathcal{L}_{M}= -(1 - \mathbf{m}_{t})^{\gamma}\log(\mathbf{m}_{t})
\end{equation}
where $\gamma \geq 0$ is a tunable focusing parameter.
Intuitively, this scaling factor can automatically down-weight the contribution of easy elements (the background and the palm areas) during training and rapidly focus the model on hard elements. In particular, if the element is misclassified and $\mathbf{m}_{t}$ is small, the factor is near 1 and the loss is unaffected. As $\mathbf{m}_{t} \to 1$, the factor goes to 0 and the loss for well-classified elements is down-weighted. The focusing parameter $\gamma$ is
increased the effect of the factor is likewise increased. When $\gamma = 0$, $\mathcal{L}_{M}$ is equivalent to $\mathcal{L}_{BCE}$. $\gamma$ is set to 2 in our case.

\subsubsection{3D Hand Pose Estimation}
Another pose regression decoder is used to infer the 3D hand pose, $\hat{\textbf{\textit{J}}} \in \mathbb{R}^{21\times3}$, from the deep representation $\textbf{r}$, with the supervision of ground truth as illustrated in the bottom right of Figure~\ref{f:handnet} such that: 
\begin{equation}
\setlength\abovedisplayskip{3pt}
\setlength\belowdisplayskip{3pt}
\mathcal{L}_{J}=\frac{1}{n} \sum_{i=1}^n(\textbf{\textit{J}} - \hat{\textbf{\textit{J}}}) 
\end{equation}
where $n$ is the number of training samples of the hand pose. When solely doing pose estimate task using the $\mathcal{L}_{J}$, we note that the network easily gets stuck at a local optimum, resulting in 21 joints clustering around a specific location and therefore the network hard to regress a meaningful hand pose. When learning concurrently with the hand mask generation task, this phenomenon disappears, implying that the hand's structure aids in regularizing hand pose reconstruction. Inspired by this observation, additional prior knowledge about the hand can be imported into the pose decoder to further constrain the estimated hand pose.
First, \networkname{} adopts bone length loss~\cite{sun2017compositional, liu2019real} that is utilized in human body estimation task to constrain the fingers by penalizing invalid bone lengths. As mentioned in Section~\ref{s:hand model}, 
hand is an articulated object comprising the palm and five independent fingers, and each of the finger consists of three types of bones arranged in a hierarchical manner. The bone length regularization loss is defined such that:
\begin{equation}
\mathcal{L}_{BL}=\frac{1}{15} \sum_{i=1}^{15} \mathcal{R}\left(\left\|\mathbf{b}_i\right\|_2 ; b_i^{\min }, b_i^{\max }\right)
\end{equation}
Here, $\mathcal{R}$ denotes a range-constrained function that restricts the variable $x$ to fall within a specific range $[a,b]$ such that:
\begin{equation}
\mathcal{R}(x ; a, b)=\max (a-x, 0)+\max (x-b, 0)
\end{equation}


Each of the bones $i$ has a range $\left[b_i^{\min }, b_i^{\max }\right]$ of valid length that can be found in~\cite{chen2013constraint}.

In addition, the five CMC bones exhibit a lower degree of freedom compared to other hand bones. The five CMC bones, along with the root joint and five CMC joints, collectively span a palmar structure~\cite{spurr2020weakly}. The four angular distance $\phi$ of five CMC bones is within a certain degree from each other in the plane they span~\cite{ryf1995neutral}. The angle is calculated by:
\begin{equation}
\phi_i= \arccos \left(\frac{\mathbf{b}_i^T \mathbf{b}_{i+1}}{\left\|\mathbf{b}_i\right\|_2\left\|\mathbf{b}_{i+1}\right\|_2}\right)
\end{equation}
At the same time, the curvature $c$ between two CMC joints is limited in a range as well. The curvature between two bones can be calculated approximately by following~\cite{rusinkiewicz2004estimating}:
\begin{equation}
c_i=\frac{\left(\mathbf{e}_{i+1}-\mathbf{e}_i\right)^T\left(\mathbf{b}_{i+1}-\mathbf{b}_i\right)}{\left\|\mathbf{b}_{i+1}-\mathbf{b}_i\right\|^2}, \text { for } i \in\{1,2,3,4\}
\end{equation}
where $\mathbf{e}_i$ is the edge norm at bone $\mathbf{b}_i$ such that:
\begin{equation}
\begin{aligned}
& \mathbf{n}_i=\operatorname{norm}\left(\mathbf{b}_{i+1} \times \mathbf{b}_i\right), \text { for } i \in\{1,2,3,4\} \\
& \mathbf{e}_i= \begin{cases}\mathbf{n}_1, & \text { if } i=1 \\
\operatorname{norm}\left(\mathbf{n}_i+\mathbf{n}_{i-1}\right), & \text { if } i \in\{2,3,4\} \\
\mathbf{n}_4, & \text { if } i=5\end{cases}
\end{aligned}
\end{equation}
where $\operatorname{norm}(\mathbf{x})=\frac{\mathbf{x}}{\|\mathbf{x}\|_2}$. To ensure biomechanical valid palm structures, both angular distance $\phi$ and curvature $c$ are taken into consideration and define the palmar structure constraints loss as follows:
\begin{equation}
\mathcal{L}_{P}=\frac{1}{4} \sum_{i=1}^4\left(\mathcal{R}\left(c_i ; c_i^{\min }, c_i^{\max }\right)+\mathcal{R}\left(\phi_i ; \phi_i^{\min }, \phi_i^{\max }\right)\right)
\end{equation}

In summary, the pose constraints loss is constructed as follows:
\begin{equation}~\label{e:lpc}
\mathcal{L}_{PC}=\beta \mathcal{L}_{J} + \gamma \mathcal{L}_{BL} + \lambda \mathcal{L}_{P}
\end{equation}
where $\beta$,$\gamma$ and $\lambda$ are weights to balance the individual loss terms. 


\subsection{Domain Generalization}\label{training}
Different hand placement positions can result in distinct multi-path profiles with different CSI data distributions, which leads the learned model to perform poorly. In fact, this issue is known as \textit{domain shift}, which refers to the change in the data distribution. In this paper, we refer to a pair of relative positions of hand and routers in a specific environment as \textit{domain}. The domain where the model is trained as source domain and the domain where the model is applied  as target domain. Literature has proposed numerous methods to cope with the domain shift issue and tried to improve the generalization of models. Traditional Empirical Risk Minimization (ERM)~\cite{vapnik1999overview} approaches aim to collect sufficient data from diverse domains to provide the model with strong generalization capabilities. Alternatively, domain adaptation~\cite{ben2010theory} approaches adjust network parameters to adapt the learned model to different target domains. However, collecting data from all domains is a costly and impractical endeavor. Domain adaptation assumes that the model has access to some information about the new domain, enabling it to perform meta-learning or few-shot learning, which is not the case for \systemname{} scenario.   When deploying \systemname{} in real-world scenarios, one of the challenges is that we may not know the exact relative location of the present hand or the WiFi routers. As such, the model must distinguish the causal signal feature of interest rather than spurious signal feature specific to a target domain. An opportunity for \systemname{} is that regardless the different domain, our label's distribution are always the same. Therefore, \systemname{} resorts to the advanced domain generalization (DG) technique.
%

In particular, suppose we have $K$ similar but distinct source domains $\mathcal{S}=\left\{S_k=\left\{\left(x^{(k)}, y^{(k)}\right)\right\}\right\}_{k=1}^K$, where $x$ is the CSI data, $y$ is the label, each associated with a joint distribution ${P}_{XY}^{(k)}$. Note that ${P}_{XY}^{(k)} \neq {P}_{XY}^{(k')}$ with $K \neq K'$ and $k,k' \in \left\{1,...,K\right\}$. Note that the marginal distribution of label space remains the same, ${P}_{Y}^{(\mathcal{T})} = {P}_{Y}^{(\mathcal{S})}$.
DG tries to learn a model $\mathcal{F}: \mathcal{X} \rightarrow \mathcal{Y}$ from several different but related source domains data such that the error of the model on  \textit{unseen} target domain $\mathcal{T}=\left\{x^{\mathcal{T}}\right\}$ is minimized~\cite{blanchard2011generalizing,9847099}. 
The assumption behind this is that even if we do not have information from all domains, it is possible to have a subset of domains that are sufficient for identifying true causal features and distinguishing them from statistically associated features that may vary across different domains~\cite{Arjovsky2019InvariantRM}.

To learn the position-agnostic features, we adopt deepCORAL~\cite{sun2016deep} into \networkname{}. Concretely, we collect the same set of data but in different domain and obtain their deep representation $S_1 = \left\{\textbf{\textit{r}}_{i}^{1}\right\}$ and $S_2 = \left\{\textbf{\textit{r}}_{i}^{2}\right\}$, where we choose two domains for clarification. 
In practice, there can be many domains.
A CORAL loss is applied to minimize the distance between the second-order statistics (covariances) of the different domain latent features:
\begin{equation}
\mathcal{L}_{C O R A L}=\frac{1}{4 d^2}\left\|\mathbf{C}_{S_1}-\mathbf{C}_{S_2}\right\|_F^2
\end{equation}
where $d$ is the dimension of $\textbf{r}$ and $\|\cdot\|_F^2$ denotes the squared matrix Frobenius norm. The covariance matrices are given by:
\begin{equation}
\begin{small}
\begin{aligned}
\mathbf{C}_{S_1} & =\frac{1}{bs-1}\left(S_1^{\top} S_1-\frac{1}{bs}\left(\mathbf{1}^{\top} S_1\right)^{\top}\left(\mathbf{1}^{\top} S_1\right)\right) \\
\mathbf{C}_{S_2} & =\frac{1}{bs-1}\left(S_2^{\top} S_2-\frac{1}{bs}\left(\mathbf{1}^{\top} S_2\right)^{\top}\left(\mathbf{1}^{\top} S_2\right)\right)
\end{aligned}
\end{small}
\end{equation}
where $bs$ is the batch size and $\mathbf{1}$ is a column vector with all elements equal to 1. Note that $\mathcal{L}_{C O R A L}$ is differentiable so it can incorporate to end-to-end deep neural network and the gradient can be back-propagated. 
Unlike the original CORAL method, which aims to align the feature representations of different networks before the classification head, \networkname{} uses DeepCORAL by alternately feeding data from different domains on a batch basis to compel \networkname{} to capture domain-invariant causal feature during the extraction of deep hand semantic information.
The whole \networkname{} is trained end-to-end with the summation of all the losses such that Eq~\ref{e:lpc} will be minimized: 
\begin{equation}~\label{e:lpc}
\mathcal{L}=\alpha \mathcal{L}_{M} + \beta \mathcal{L}_{J} + \gamma \mathcal{L}_{BL} + \lambda \mathcal{L}_{P} + \zeta
\mathcal{L}_{C O R A L}
\end{equation}
where $\alpha,\beta,\gamma,\lambda$ and $\zeta $ are weights to balance the individual loss
terms.


\section{Evaluation}
\subsection{Implementation}
\paragraph{Commercial Wi-Fi:} 
We collect the CSI data using two commercial routers (TP-LINK WDR4310) equipped with Atheros SoC AR9344. The transmitting router has a single antenna, while the receiving router has three antennas. The raw CSI data is collected using the Atheros-CSI-Tool~\cite{xie2015precise} and sent to a server through a TCP link using the general socket API. The WiFi signals are transmitted on the 5GHz channel with a 40MHz bandwidth, and the packet rate is set at 200 packets per second.

\paragraph{Server:}
The server employed for both CSI data collection and model inference is a Linux desktop computer equipped with an Intel Core i9-9820X CPU and one Nvidia 2080Ti GPU card.

\paragraph{Leap Motion:}
The ground truth 2D mask and 3D joint data are obtained using a commercial egocentric depth infrared stereo camera, Leap Motion. The depth camera provides sub-millimeter accuracy of the hand joints within a field of view of approximately $135^{\circ}$, with an  effective sensing range above the device of approximately 25 to 600 millimeters~\cite{leap}.

\paragraph{Ground truths:}
The Leap Motion sensor is positioned at 30cm directly below the hand. Using the official python API, we collect the 3D coordinates of 21 hand joints as defined by our hand model as our ground truth. We then normalize the coordinates from 0 to 1. A 2D hand mask is generated from the leap image using the GrabCut~\cite{10.1145/1186562.1015720} algorithm by applying OpenCV for foreground segmentation to extract the hand from the background. The resulting mask uses a binary representation, with 1 representing the hand area and 0 representing the background. To avoid the class imbalance problem, we crop the hand mask with the reference of the center of the palm. We use a local network time protocal (NTP) server to synchronize the routers and the depth camera.
\begin{figure}[h]
  \centering
  \includegraphics[width=0.7\linewidth]{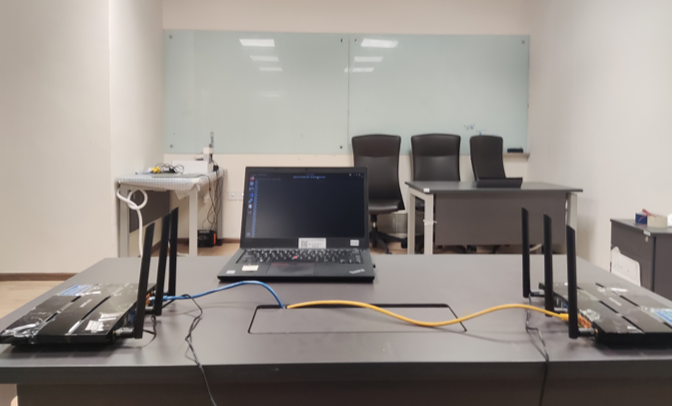}
  \caption{Testbed.}
  \label{f:testbed}
\end{figure}

\paragraph{Data Collection: }
We invited 6 subjects and collected data in 5 different environments with 5 different 
hand positions relative to the WiFi transceivers. \footnote{All experiments that involve humans have been approved by IRB.} In order to validate the effectiveness of our system on domain shifts (e.g., unseen subjects, gestures, environments, and hand positions), we construct a training dataset (main dataset) from two subjects for model training. 
The training dataset includes 27 hand postures consisting of 26 American Sign Language (ASL) alphabets and one fully open hand posture. As suggested by RFPose~\cite{zhao2018through}, the specularity of the human body causes unequal reflection, and therefore we collect data as a sequence to enhance the learning process.
Each posture is performed within a 2-second window with a slight up-and-down motion, except for the dynamic alphabets J and Z. This approach is also consistent with the principles for super-resolution, which involves utilizing complementary information from adjacent frames to perform super-resolution reconstruction~\cite{tian2011survey}. Deep neural networks are a powerful tool for achieving super-resolution by learning patterns and features from a large amount of data~\cite{yang2019deep}.
The training dataset is collected in a testbed as shown in Figure~\ref{f:testbed}, where the leap motion is placed at the center of the two WiFi routers. Volunteers are invited to perform 27 hand gestures between the routers and above the leap motion. 
Two volunteers (\#P1: 21-year old male and \#P2: 27-year old female) perform each gesture for about 4.5 minutes respectively. As the router outputs 200 CSI per second and we consider every 20 CSI measurements as one sample, we generate 10 samples per second. Thus, each sample can be represented as a vector with the size of 3x114x20, representing 3 spatial streams, 114 WiFi subcarriers, and 20 CSI samples, respectively. The training dataset is randomly split into 9:1 for training and validation of the HandNet model.

\paragraph{Training details:}
The HandNet is trained using the ADAM optimizer~\cite{kingma2014adam} with Cosine Annealing Learning Rate scheduler~\cite{loshchilovsgdr}. The initial learning rate is 0.001. The batch size is 24 and HandNet is implemented by PyTorch.

\subsection{Evaluation Metric}

Figure~\ref{f:keyfigure} shows the qualitative results of \systemname{}, which demonstrates that the WiFi-based \systemname{} is capable of achieving competitive perceptual accuracy to that of the CV-based system. 
To further analyze it, we adopt several metrics that are widely measured in the CV domain. The mean Pixel Accuracy (mPA) and Intersection-over-Union (IoU) are used for 2D mask evaluation. The Mean Per Joint Position Error (MPJPE) and the Percentage of Correct Keypoints (PCK) are used for 3D joint evaluation. The metrics are detailed below:

\paragraph{mPA:}
Pixel accuracy is a classic semantic segmentation metric, defined as
the ratio between the amount of accurately classified pixels and the total number of pixels in the image. Accordingly, mPA represents the average percentage of accurately classified pixels in the entire image.:
\begin{equation}
mPA=\frac{1}{k} \sum_{i=0}^{k}\frac{p_{ii}}{\sum_{j=0}^{k}p_{ij}}
\end{equation}
where $p_{ii}$ is the total number of true positive pixels for class $i$, $p_{ij}$ stands for the pixel that belongs to class $i$ but mistakenly as class $j$ and $k$ is the total number of class, where $k=2$ in our case.
A drawback of mPA is that it favors the dominant class. For example, if the hand is very far away from the camera and just occupies a small portion of the mask, the background class is dominant. In an extreme case,  even if a predicted mask is without a hand at all,  the accuracy will still be almost 100\%. Thus, we introduce IoU as well.

\paragraph{IoU:}
IoU is the area of overlap between the predicted hand and the ground truth hand in the mask, divided by the area of union between the predicted hand and the ground truth hand:
\begin{equation}
IoU=\frac{p_{ii}}{p_{ii}+p_{ji}+p_{ij}}
\end{equation}
\paragraph{MPJPE:}
MPJPE calculates the average Euclidean distance between the predicted joint coordinates and the ground truth joint coordinates:
\begin{equation}
MPJPE=\frac{\sum_{i=1}^{J}\left( {\left\|\mathbf{\hat{j}}_{i}-\mathbf{j}_i\right\|}_{2}\right)}{J}
\end{equation}

\paragraph{PCK:}
PCK describes the percentage of correct predicted joints, where the predicted joint is considered correct if the distance between it and the ground truth joint is within a certain threshold:
\begin{equation}
PCK@a=\frac{\sum_{i=1}^{J} \delta\left( {\left\|\mathbf{\hat{j}}_{i}-\mathbf{j}_i\right\|_{2}  \leq a  }\right)}{J}
\end{equation}
where $\delta$ is a logical operation that outputs 1 if Ture and output 0 if False. We set $a=2cm$.

\subsection{Ablation Study}
Since there is currently no model available that can extract hand masks and poses from RF signals, we conduct an ablation study by designing several baseline (BL) models to understand the role of different components in the proposed \networkname{} architecture. Specifically, these baseline models are designed by removing or replacing specific modules of \networkname{} as follows:

\paragraph{Baseline A - single task:} This baseline only contains the multi-scale encoder of HandNet and the pose decoder with joint regression loss $\mathcal{L}_{J}$.

\paragraph{Baseline B - ResNet50 encoder:} Baseline B changes the proposed multi-scale encoder of baseline A to ResNet50~\cite{he2016deep} as the encoder.

\paragraph{Baseline C - UNet encoder:} This baseline uses UNet~\cite{ronneberger2015u} with joint regression loss $\mathcal{L}_{J}$.

\paragraph{Baseline D - signal embedding:} RF signal embedding layer + Baseline A.

\paragraph{Baseline E - multi-task:} Baseline E adds one mask decoder with MSE loss to the baseline D to make it a multi-task encoder-decoder structure.

\paragraph{Baseline F - BCE loss:} Use BCE loss to replace baseline E's MSE loss.

\paragraph{Baseline G - mask loss:} Use $\mathcal{L}_{M}$ to replace baseline F's BCE loss.

\paragraph{Baseline H - HandNet:} $\mathcal{L}_{PC}$ + baseline G, which is the whole HandNet.


\begin{table}
\centering
\caption{Ablation study of \systemname{}.}
\label{t:qr}
\begin{tabular}{l|cccc} 
\hline\hline
                 & mPA           & IoU           & MPJPE(cm)              & PCK@2cm                 \\ 
\hline\hline
BL.A-single task       & /             & /             & 20.41                  & 0.19                    \\
BL.B-ResNet50       & /             & /             & 25.27                  & 0.10                    \\
BL.C-UNet       & /             & /             & 22.65                  & 0.14                    \\ 
\hline
BL.D-embedding       & /             & /             & 13.98                  & 0.29                    \\ 
\hline
BL.E-multi-task       & 0.90          & 0.78          & \textit{6.43}          & \textit{0.71}           \\ 
\hline
BL.F-BCE loss       & 0.92          & 0.80          & 6.41                   & 0.72                    \\
BL.G-mask loss       & 0.94          & 0.91          & 5.72                   & 0.76                    \\
\textbf{BL.H-HandNet} & \textbf{0.94} & \textbf{0.91} & \textbf{\textit{2.07}} & \textbf{\textit{0.93}} 
\end{tabular}
\end{table}

Table~\ref{t:qr} presents the average quantitative results for 5 runs of each baseline on the main training dataset. The aim of Baseline A-C is to evaluate different choices of backbone networks. ResNet50~\cite{he2016deep} and  UNet~\cite{ronneberger2015u} are popular backbones with good pose estimation performance in CV domain. While UNet is better than ResNet in terms of lower average joint error (MPJPE) and higher correct joint rate (PCK), our proposed multi-scale backbone outperforms them both. The reason for this is that our multi-scale backbone focuses on extracting features from different scales, instead of solely focusing on identifying useful features. 

Baselines D and E evaluate the necessity of the RF signal embedding layer and the multi-task structure. The results show that organizing the complex-valued RF signal properly boosts the accuracy by about 42.07\%, and the added multi-task structure takes the hand pose accuracy to another level. 
Baselines F-H evaluates different loss functions, with the BCE loss generally providing better results than the MSE loss. The $\mathcal{L}_{M}$ significantly increases the IoU result by improving the finger boundary while also benefiting the pose estimation task. The $\mathcal{L}_{PC}$ loss provides an additional lift for the hand pose accuracy. The added mask generation task and the designed constraints for hand joints are key to the success of HandFi.

\subsection{Evaluation of Robustness}
We test \systemname{} in various conditions to evaluate the robustness and the characteristics of the system.

\subsubsection{Under Occlusion}
\begin{figure}[t]
	\begin{minipage}[t]{0.95\linewidth}
    
		\centering
		\includegraphics[]{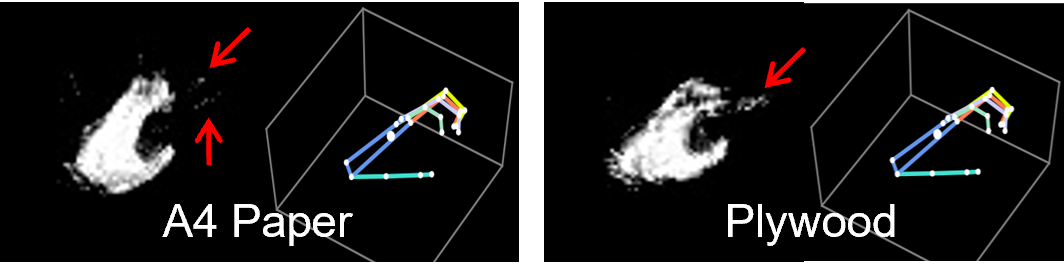}
	\end{minipage}
	\begin{minipage}[t]{\linewidth}
		\centering
  \small
	\begin{tabular}{c|cc|cc} 

\hline\hline
Scenarios   & \multicolumn{2}{c|}{A4 Paper} & \multicolumn{2}{c}{Plywood}  \\ 
\hline\hline
Leap Motion & \multicolumn{2}{c|}{X}        & \multicolumn{2}{c}{X}        \\ 
\hline
HandFi      & IoU:0.78 & PCK:0.90                  & IoU:0.67 & PCK:0.86                 \\
\hline
\end{tabular}	
\caption{Examples of masks and poses under occluded
scenarios and \systemname{} accuracy under occlusion.}
\label{t:occ}
	\end{minipage}
 \label{f:occ}
\end{figure}

Another advantage of RF sensing is that it can sense under occlusion. We use an A4 paper and a plywood to occlude the hand and play the ASL letter c 10 times independently and let the \systemname{} infers the hand information. We note that because of the occlusion, the Leap Motion cannot work in this experiment. The quantitative result\footnote{The ground truth is obtained by removing the occlusion immediately.} is reported in the table of Figure~\ref{t:occ} along with examples of the hand mask and hand pose. As expected, \systemname{} works well under occlusion, but interestingly, we saw some extra components on the masks due to the reflection of the occlusion.


\subsubsection{Effective Sensing Range}
WiFi offers omnidirectional ($360^{\circ}$) sensing as demonstrated in \S~\ref{360}, while Leap Motion has the field of view of a cone roughly $135^{\circ}$. The sensing distance of Leap Motion is limited by its FoV and can be roughly calculated by $P*tan(135/2)$, where $P$ indicates the perpendicular distance of the hand to Leap Motion. For example, if the hand is placed 30cm above the Leap Motion, the sensing range for Leap Motion is +/- 23cm. In contrast, WiFi has a room-scale sensing range. In the following, we conduct the experiment to understand the effective sensing range of \systemname{}. 

\begin{figure}[h!]
  \centering
  \includegraphics[width=\linewidth]{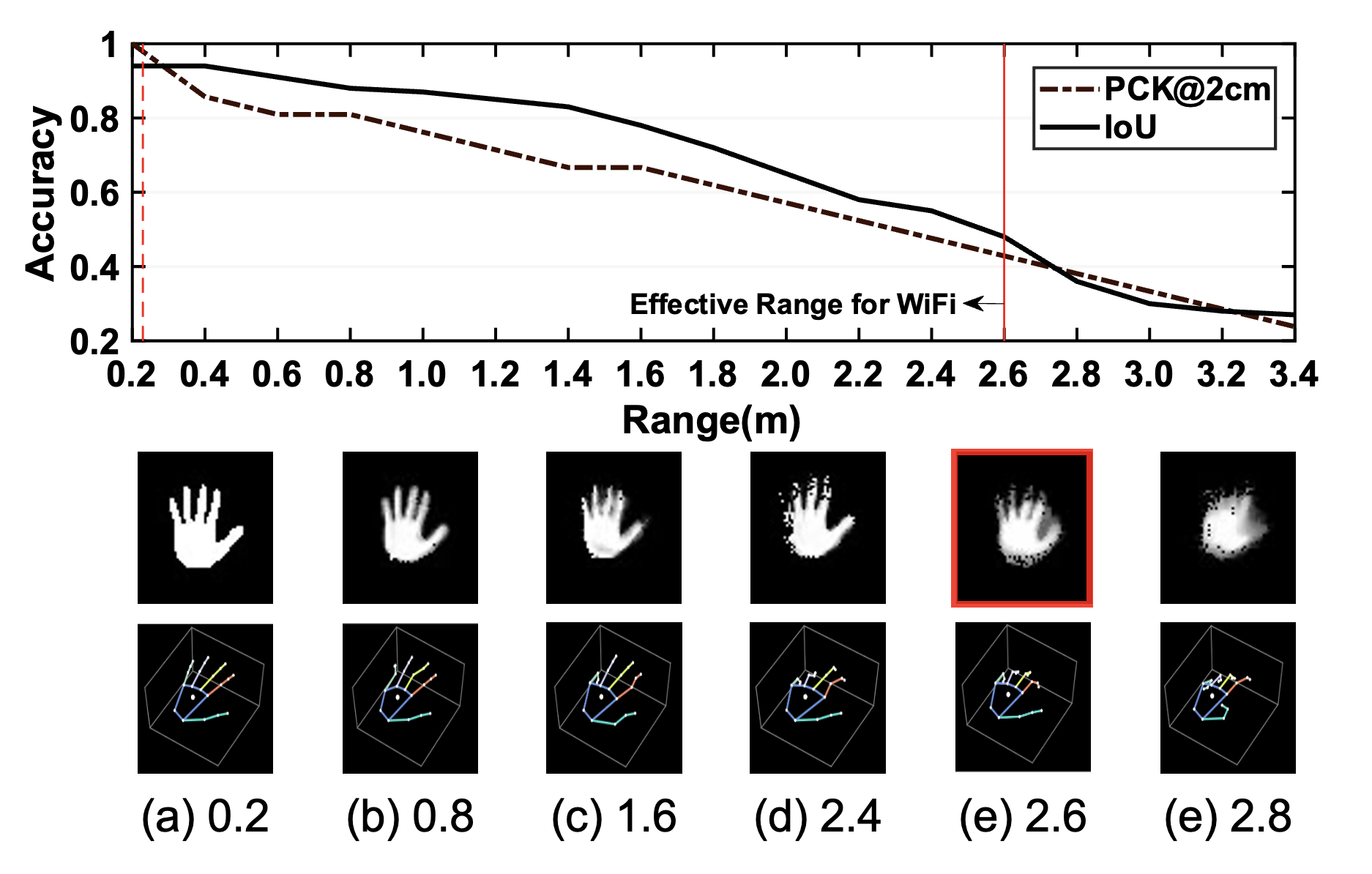}
  \caption{Quantitative results of \systemname{} at different sensing distances and the inferred hand mask at the corresponding sensing distances.}
\vspace{-0.3cm}
  \label{f:rangexp}
\end{figure}

In particular, starting from the midpoint between two WiFi APs, we place an open hand at every 0.2-meter interval. The hand holds roughly at the same plane level, and at each range, we collect 1s with 10 samples for one-time hand placement. We compute the IoU of the mask output by the HandFi system and the PCK@2cm value of the pose at each range. Figure~\ref{f:rangexp} reports the results. The results are presented in Figure 10, where we showcase the inferred 2D hand masks at distances of 0.2m, 0.8m, 1.6m, 2.4m, and 2.6m. It can be observed from the masks that their clarity decreases with increasing distance. Some hand areas are mistakenly identified as background areas, resulting in holes in the masks, as can be seen in mask (b). Moreover, misclassifications of finger areas become increasingly frequent, as evident from masks (c) and (d). Finally, the model becomes more and more confused with other gestures, as demonstrated by mask (e). As for the pose, although the accuracy of finger joints is getting worse, the palm structure still aligns with the open palm. 
Since at 2.6m, we are still able to tell that the mask/pose is an open hand and thus the effective sensing range of WiFi is +/- 2.6m, x11 larger than Leap Motion.

\subsubsection{Different Environments}
\begin{figure}[t]
	\begin{minipage}[t]{\linewidth}
   \centering 
\begin{subfigure}{0.43\linewidth}
  \includegraphics[width=\linewidth]{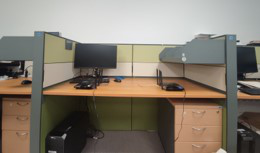}
  \caption{Env. 2: Office cubicle. }
  \label{fig:cub}
\end{subfigure}\hfil 
\begin{subfigure}{0.43\linewidth}
  \includegraphics[width=\linewidth]{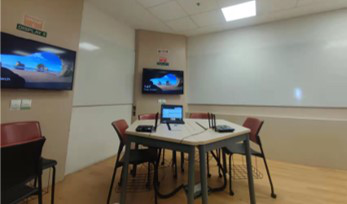}
  \caption{Env. 3: Tutorial room.}
  \label{fig:lec}
\end{subfigure}\hfil 

\begin{subfigure}{0.43\linewidth}
  \includegraphics[width=\linewidth]{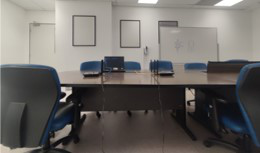}
  \caption{Env. 4: Meeting room.}
  \label{fig:4}
\end{subfigure}\hfil 
\begin{subfigure}{0.43\linewidth}
  \includegraphics[width=\linewidth]{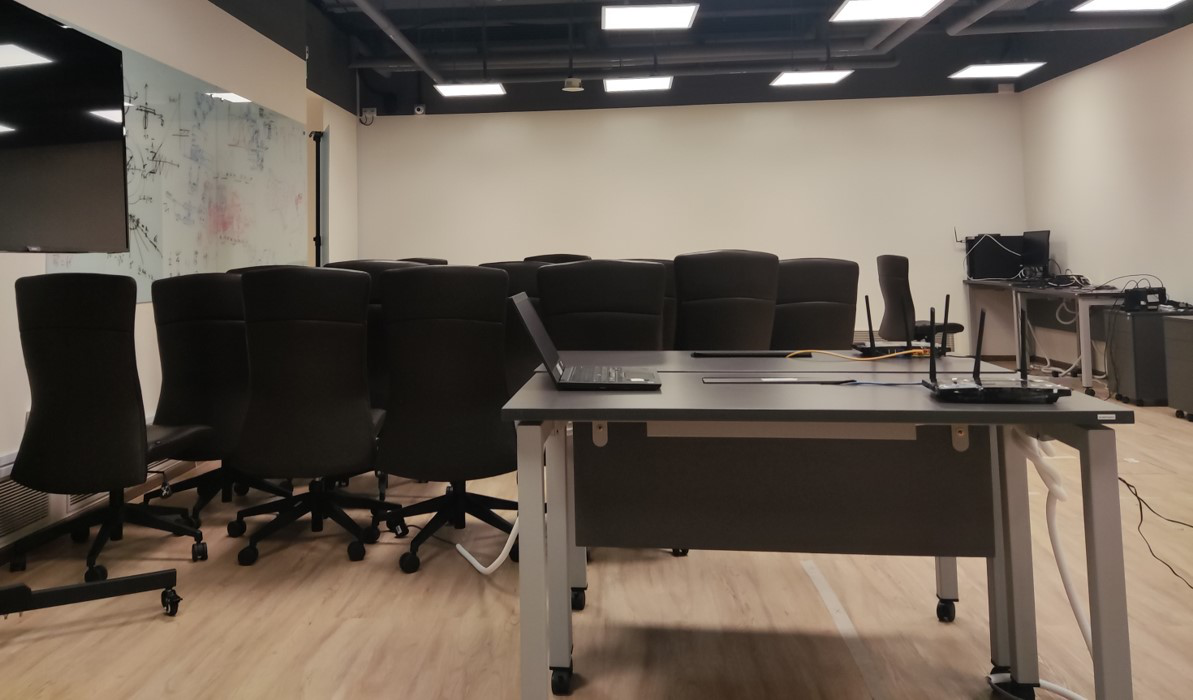}
  \caption{Env. 5: Project room.}
  \label{fig:5}
\end{subfigure}\hfil 
	\end{minipage}
	\begin{minipage}[t]{\linewidth}
		\centering
	 \includegraphics[width=\linewidth]{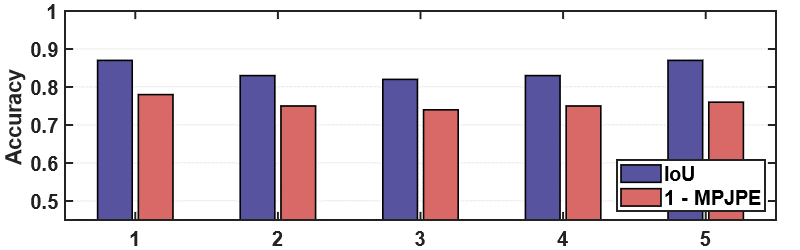}
 \caption{Performance of \systemname{} under  different environments.}
\vspace{-0.3cm}
  \label{f:envs}
	\end{minipage}
 
\end{figure}

We directly migrate \systemname{} that was trained in the testbed (Fig~\ref{f:testbed}) to several typical indoor environments as illustrated in Figure~\ref{f:envs}, and ask the subject \#P1 to perform the rock-scissors-paper gesture, where the rock and scissors gestures were not seen in the training. Each gesture is performed by ten times. The average IoU and 1-MPJPE are reported in Figure~\ref{f:envs}, and the mean variances of the mask and pose results are 0.058\% and 0.023\%. Therefore it is safe to say \systemname{} is robust to different environments.

\begin{table}[h]
\centering
\caption{Performance of \systemname{} under different users (gender and age).}
\begin{tabular}{ccc|cccc} 
\hline\hline
      & M,21(\#P1) & F,27({\#P2}) & M,22 & M,20 & M,21 & F,20  \\ 
\hline\hline
IoU   & 0.91 & 0.88 & 0.84 & 0.85 & 0.85 & 0.81  \\ 
\hline
MPJPE & 1.95 & 2.07 & 2.24 & 2.19 & 2.21 & 2.71  \\
\hline
\end{tabular}
\label{t:mu}
\end{table}

\subsubsection{Different Users}
We conducted a study involving 6 subjects of varying ages and genders, who were instructed to perform rock-scissors-paper gestures. Each gesture was repeated ten times for each subject in our Testbed 1.
The results are reported in Table~\ref{t:mu}.
Note that the first and the second users contribute to the training dataset of \systemname{}. 
The results suggest that the accuracy of \systemname{} decreases slightly for unseen users, but still outperforms the baselines. In addition, we observed that the system performs less accurately for female users compared to male users. This could be because female hands are generally smaller and have less reflection information. We leave the investigation of this observation for our future work.

\subsection{Different Location of the Hand}\label{360}
\begin{figure}[h]
	\begin{minipage}[t]{0.34\linewidth}
		\centering
		\includegraphics[width=\linewidth]{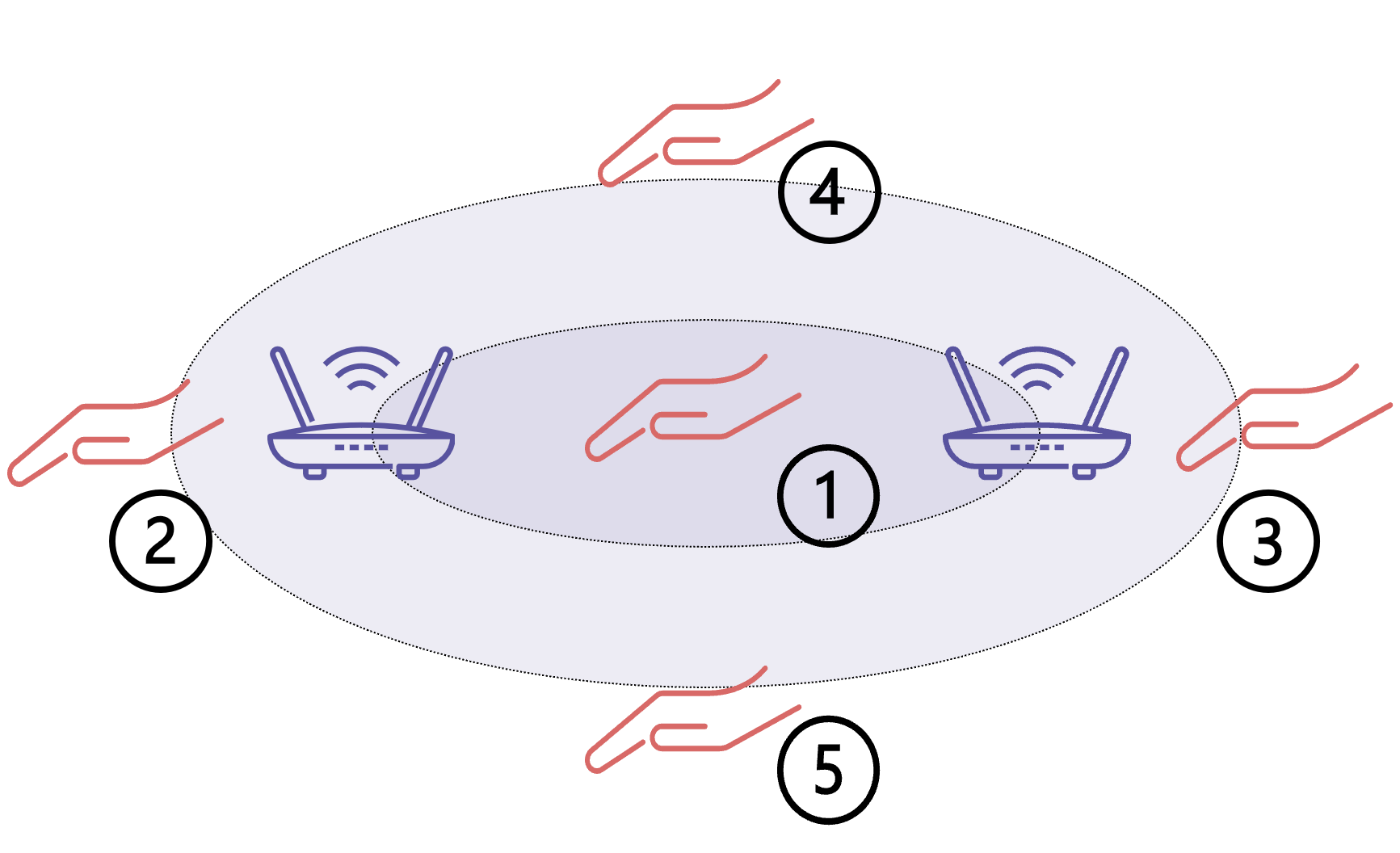}
		\caption{Hand \\ positions .}
		\label{f:pos}
	\end{minipage}
	\begin{minipage}[t]{0.65\linewidth}
		\centering
		\includegraphics[width=\linewidth]{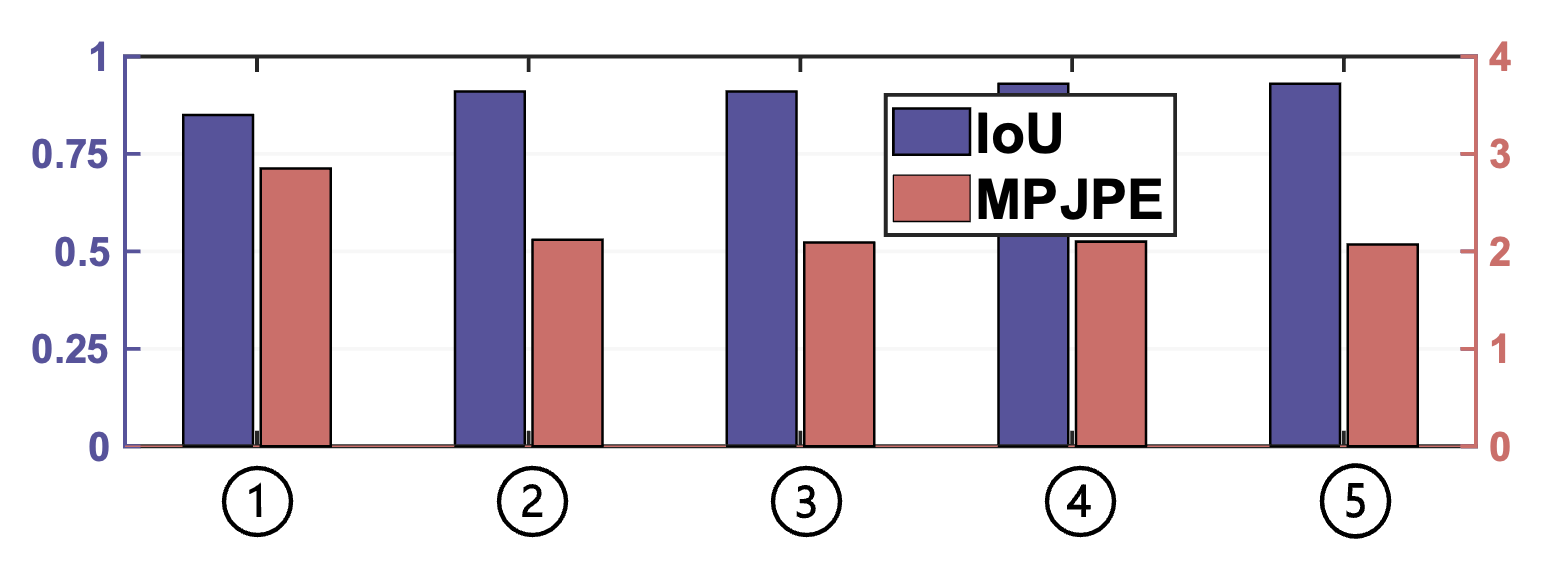}
		\caption{Performance of \systemname{} in different hand positions with domain generalization employed (trained on other 4 positions).}
		\label{f:dc}
	\end{minipage}
\end{figure}

\begin{figure}[h]
    \centering 
\begin{subfigure}{0.49\linewidth}
 \includegraphics[height=2.6cm,width=\linewidth]{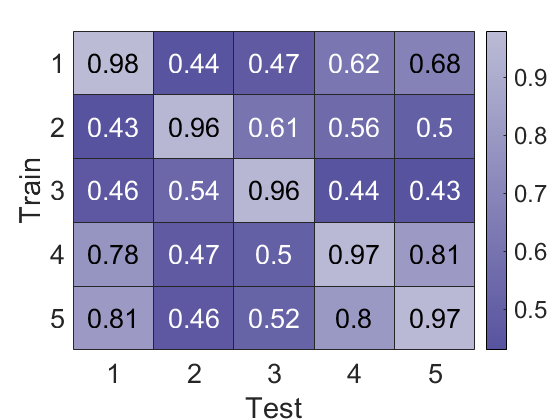}
		\caption{IoU}
		\label{f:IoU}
\end{subfigure} 
\begin{subfigure}{0.49\linewidth}
  \includegraphics[height=2.6cm,width=\linewidth]{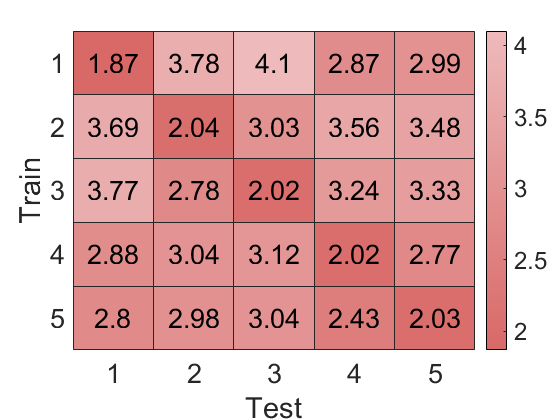}
  \caption{MPJPE(cm)}
  \label{f:MPJPE}
\end{subfigure} 
\caption{Performance of \systemname{} in different hand positions without domain generalization employed.}
\label{f:posr}
\end{figure}

To evaluate the effectiveness of the domain generalization technique, we collected datasets at different hand positions and conducted evaluations. Specifically, we collected data using Testbed 1 (Figure~\ref{f:testbed}) in Env. 5 (Figure~\ref{fig:5}). The volunteers performed hand gestures at 5 different positions (domains) as illustrated in Figure~\ref{f:pos}
The data collection process at each hand position is the same as that of the main dataset (Section 4.1). Data from the same position are aggregated into a dataset, and we collected 5 datasets at different positions in total.
We report the quantitative results in Figure~\ref{f:posr} when  \systemname{} is trained and tested in the same dataset as well as different datasets. 
As can be seen from Figure~\ref{f:posr}, when \systemname{} is tested on the same domain (i.e., the diagonal line of the table), its performance is consistently good regardless of the position of the hand with ~0.97 IoU of the hand mask and ~2.00cm joint errors. However, when testing \systemname{} on different domains (i.e., the non-diagonal part of the table), a significant performance drop is observed, with a range of 16.5-56.1\% for the mask task and 20.3-103.0\% for the pose task, respectively.   


By adopting the domain generalization training strategy introduced in Section~\ref{training}, we train the model with different dataset from 4 different positions and evaluate \systemname{} in the unseen hand position and report the quantitative results in Figure~\ref{f:dc}. Compared to Figure~\ref{f:IoU} and Figure~\ref{f:MPJPE}, although neither IoU nor MPJPE results can achieve the reconstructed mask and pose as well as the diagonal result in Figure~\ref{f:posr}, the performance drop in the unseen domain is effectively mitigated, and comparable results are achieved by using the domain generalization training strategy.

\subsection{Discussion}
The current version of HandFi operates with a fixed transceiver setting around one meter apart to ensure the quality of WiFi signals. In our experiments, we observe that when the distance between transceivers increases, our system performance could be affected. We are investigating whether the performance degradation is mainly due to signal attenuation or the limitation of our model generalizability. 
In practice, we believe it can be acceptable for users to configure transceivers so as to ensure signal strength and high sensing accuracy.


HandFi is currently designed to work with one hand of a user at a time. This restricts HandFi from supporting tasks that require the coordination of both hands or involve more complex multi-user interactions. We are investigating if data augmentation techniques can be applied to generalize the model to both hands without collecting an excessive amount of new data. Multi-user scenario is challenging and we plan to explore if multi-antenna could help separate reflections from multiple users. We leave the above research problems for our future work. 

\section{Application Cases}
We developed two downstream applications on top of \systemname{} to show that the obtained 2D hand mask and 3D hand pose can directly boost the hand-related applications.
\subsection{Sign Language Recognition}
\begin{figure}[t]
  \centering
  \includegraphics[width=\linewidth]{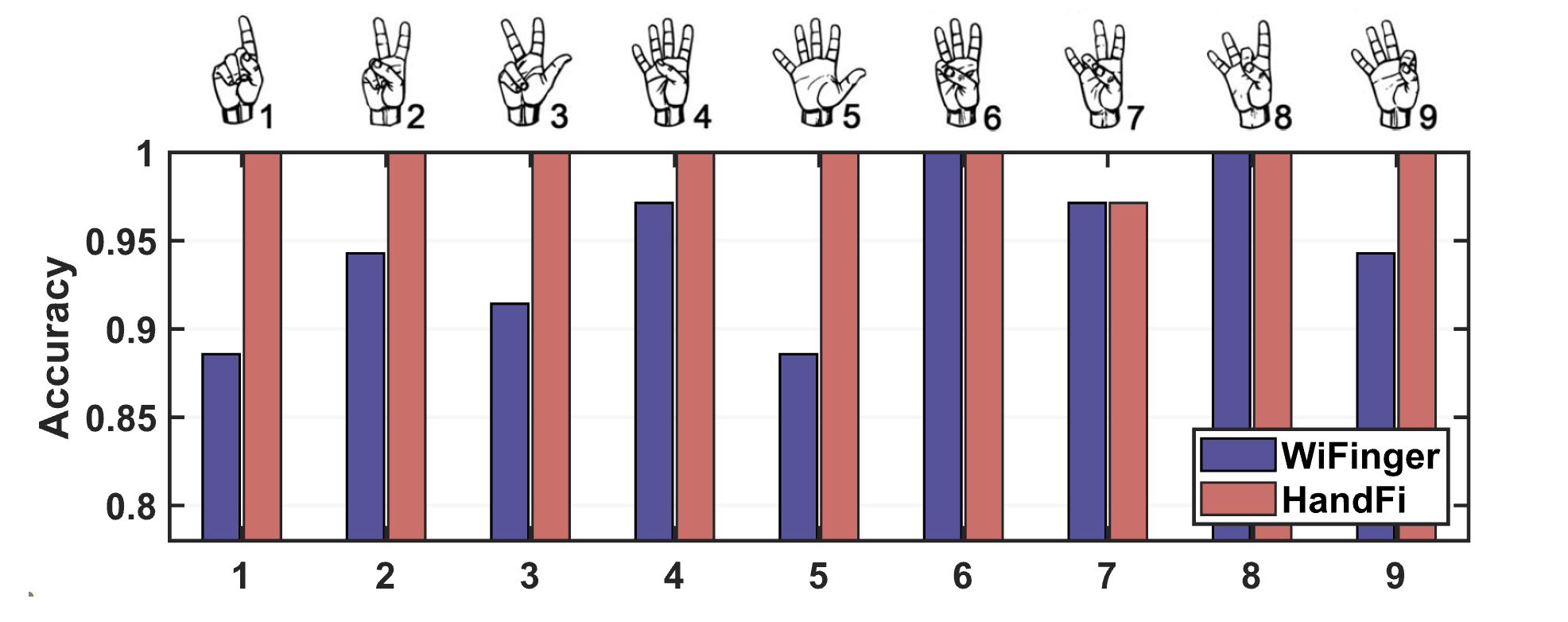}
  \caption{Break down recognition accuracy of american sign language (ASL) digits.}
  \label{f:sl}
\end{figure}

Sign language recognition is the most challenging task because of the inter-similarity between different signs and the fine-grained finger motion. WiFinger~\cite{li2016wifinger} is the first solution to use commercial WiFi to recognize 9-digit finger-level signs from American Sign Language (ASL), the most widely used sign language. WiFinger only uses one pair of the transmitter as \systemname{} does and it proposes a series of signal processing techniques with the k-Nearest Neighbor (KNN) and Dynamic Time Wrapping (DTW) to classify 9 sign digits. We conducted the same experiment as WiFinger, collecting the same amount of data from a single user, with each of the 9 sign digits having 35 samples. The single fully connected layer with softmax function serves as the classification head plug after the \networkname{} to classify the 9 digits. The results are reported in Figure~\ref{f:sl}. We see that except for number 7, all of the sign digits are classified with 100\% accuracy with the benefit of clear semantic hand information from the generated mask. Compared to the 94.60\% accuracy achieved by the WiFinger, \systemname{} achieves 99.68\% accuracy with 5.37\% performance increase. 
It is worth noting that these 9 gestures are unseen gestures for \systemname{}. \systemname{} is able to reconstruct the mask with IoU = 0.81 and the pose with MPJPE = 4.37cm and PCK@2cm = 0.78. It can be envisioned that once the low-level features of the hand are extracted, gesture recognition becomes easier and more stable.

\subsection{Finger Tracking}
Finger tracking, or finger drawing, offers natural and fine-grained interaction with hardware devices. Although it is well-studied in the literature, WiFi-based finger tracking is extremely challenging because of the low resolution of CSI. 
Since \systemname{} is able to obtain 3D hand pose, finger tracking becomes a relatively easy task, which focuses on the specific finger (index finger) in a clear way without worrying about self-occlusion. We conduct a finger tracking experiment following the most recent and the first sub-wavelength level finger tracking system, FingerDraw~\cite{wu2020fingerdraw}, and compare results with them. In particular, we print three templates (a triangle, the letter ‘Z’, and a closed half-circle-like letter 'D') on cardboard and use the index finger to follow the template trajectories, the same as FingerDraw. 
We collect 150 finger drawings in total. Each template has 50 drawings. The drawing speed is roughly 5 cm per second. 

\begin{table}[h]
\centering
\caption{Finger tracking accuracy.}
\label{t:ft}
\begin{tabular}{c|ccc|ccc} 
\hline\hline
                   & \multicolumn{3}{c|}{50\% error (cm)}          & \multicolumn{3}{c}{90\% error (cm)}            \\ 
\hline\hline
Methods            & $\triangle$   & Z             & D             & $\triangle$   & Z             & D              \\ 
\hline
FingerDraw         & 1.12          & 1.46          & 1.29          & 2.98          & 3.38          & 3.52           \\ 
\hline
HandFi             & \textbf{0.98} & \textbf{0.81} & \textbf{1.07} & \textbf{1.10} & \textbf{1.01} & \textbf{1.22}  \\ 
\hline
HandFi-3D &   2.20            &    1.05           &    1.50           &    2.24           &    1.15           &     1.50           \\
\hline
\end{tabular}

\end{table}

\begin{figure}[t!]
    \centering
 \includegraphics[width=\linewidth]{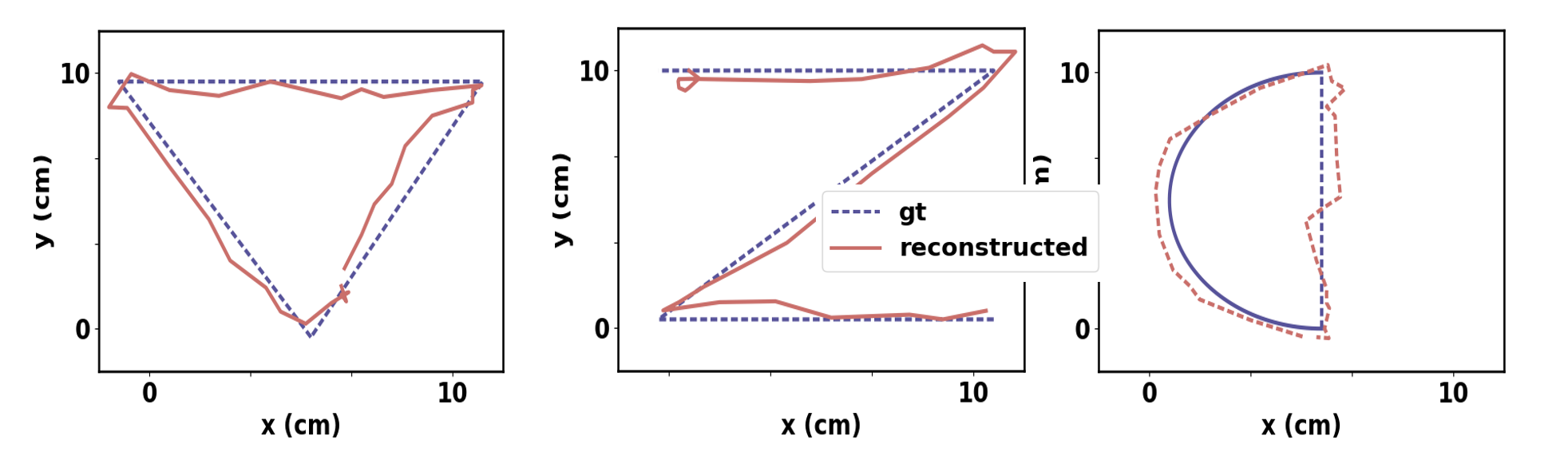}
    \caption{2D finger tracking with constraint.}
    \label{fig:2d}
\end{figure}

\begin{figure}[t]
    \centering
\includegraphics[height=2.4cm,width=\linewidth]{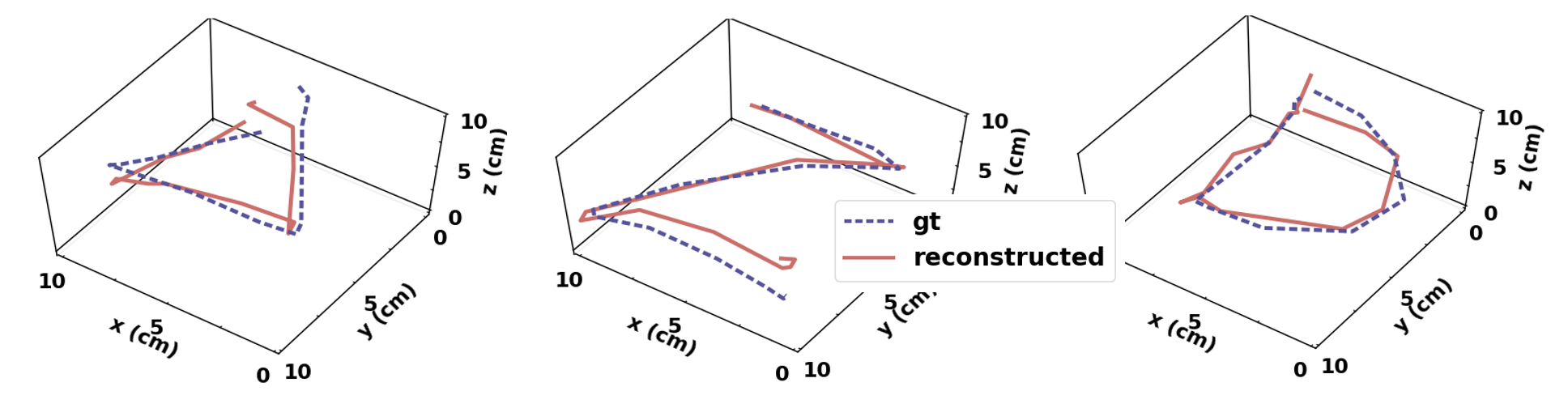}
    \caption{3D finger tracking in free space.}
    \label{fig:3d}
\end{figure}

Table~\ref{t:ft} reports the results, with \systemname{} achieving an overall median error of 0.95 cm and a 90th percentile error of 1.11 cm. These values are 24.69\% and 66.18\% higher, respectively, than those achieved by FingerDraw. Additionally, we take the cardboard off and draw the same 150 samples freely in the air. The corresponding results are reported in the last row of Table~\ref{t:ft}. Figure~\ref{fig:3d} reveals the trajectory in a free space without the constraint of cardboard, where natural jitters are observed in our finger compared to Figure~\ref{fig:2d}. This indicates HandFi's ability to capture hand tremors, which can enable health monitoring applications such as indicative of Parkinson’s disease or anxiety disorders.

In addition to \systemnames{} higher accuracy and tracking in 3D space, it is worth noting that FingerDraw has accumulative tracking errors, while \systemname{} does not have such an issue. In addition, FingerDraw requires multiple routers and at least two receivers placed orthogonality, while \systemname{} only needs one pair of transmitters.

\subsection{Computation Overhead}
Our system is evaluated in an offline manner to facilitate a more comprehensive evaluation process, but it can support online end-to-end execution. To demonstrate \systemnames{} capabilities, real-time finger tracking is performed using the trained model. The current version of \systemname{} does not solve the continuous segmentation problem. Instead, we use a sliding window plus a stop sign to conduct hard segmentation. We average the computation cost from all samples and report the computation overhead in Table~\ref{t:overhead}. The inference task can be executed on a server without a high-end GPU. The latency of CPU execution is approximately five times longer compared to GPU execution. 

\begin{table}[h!]
\centering
 \caption{Computation overhead of \systemname{}.}
   \begin{tabular}{lr}
    \hline
    \textbf{Transmission Latency (ms)} & \textbf{0.314} \\ \hline
    \textbf{Collecting Latency (s)}    & \textbf{0.1} \\ \hline
    \textbf{Inference Time (GPU) (ms)}  & \textbf{11.32} \\ \hline
    \textbf{Inference Time (CPU) (ms)}     & \textbf{50.28}  \\ \hline
    \end{tabular}
    \label{t:overhead}
    \end{table}
    

\section{related work}


\paragraph{CV-based Hand Pose Estimation.}
In the following, we review recent advanced deep learning-based hand sensing methods~\cite{doosti2019hand}. Generally, existing CV-based methods are sensitive to changes in illumination and background.
The closest works to our 2D hand mask and 3D hand pose tasks are Hand3D~\cite{zimmermann2017learning}
and HIU~\cite{zhang2021hand}, respectively. However, due to the inherent differences between WiFi signals and image signals, existing CV-based methods cannot be directly applied. Specifically, image signals inherently contain clear semantic information, with the task focusing on accurately locating the key points of hands in the image. In contrast, WiFi sensing focuses on effectively extracting features from abstract reflected signals and modeling them according to hand models. 

\paragraph{RF-based Body Pose Estimation.}
Previous RF-based human pose estimation works either require specialized FMCW devices like RF-Pose~\cite{zhao2018through} and RF-Pose3D~\cite{zhao2018rf}, or they are not robust to environmental variations~\cite{wang2019person} and require specific multi-device and multi-antenna placement~\cite{jiang2020towards}. These works all rely on body model so cannot be directly applied to hand pose estimation. Hand model has higher degrees of freedom and the reflection scale of hand is much smaller~\cite{6909541,MANO:SIGGRAPHASIA:2017}. \systemname{} is the first to use commercial WiFi to obtain vision-like hand shape and hand pose.


\paragraph{WiFi-based Hand Sensing.} Current WiFi-based hand sensing applications are built towards specific applications such as gesture recognition~\cite{abdelnasser2015wigest,pu2013whole,zhang2021widar3,tan2016wifinger}, sign language recognition~\cite{shang2017robust,zhang2020wisign,ma2018signfi,xing2022wifine}, finger tracking~\cite{tan2020enabling,zhang2018letfi,yu2018qgesture,wu2020fingerdraw}, and keystroke detection~\cite{ali2015keystroke}. In contrast, \systemname{} has the capability to support diverse applications. Furthermore, existing systems mainly rely on mapping hand motion patterns (which requires capturing a time sequence of sensing data and learning the timed pattern) to a limited set of specific gestures, rather than recognizing the basic elements of hand motion as HandFi does. Additionally, HandFi outperforms existing methods in both finger tracking~\cite{wu2020fingerdraw} and sign language recognition~\cite{li2016wifinger} tasks. \systemname{} is the first achieving 3D finger tracking.

\paragraph{Generalization in RF Sensing.}
Over-reliance on deep learning methods during training can result in performance degradation of machine learning systems when tested on data outside the training domain, commonly known as the domain shift. The heterogeneous and imperfect nature of RF data, owing to hardware imperfections and multi-path effects, exacerbates this problem. Prior research in RF sensing has sought to address this issue~\cite{ding2020rf,xiao2021onefi,yang2022environment,jiang2018towards,han2023mmsign,yangxg}, with many studies resorting to meta-learning or few-shot learning. However, these methods assume access to samples in the target domain, which may not be practical. In contrast, \systemname{} assumes the target domain is inaccessible and uses DG techniques to train a robust model with slightly lower accuracy in the source domain.

\section{Conclusion}
This paper presents \systemname{}, a WiFi hand sensing system that can construct hand shape and hand skeleton from commercial WiFi. To this end, we propose a novel symbolically constrained multi-task learning framework, \networkname{}, and incorporate the domain generalization technique to enhance the sensing performance in unseen environments. We build a prototype system and conduct comprehensive evaluation in various experiment settings. We further develop two finger-level downstream applications to demonstrate the effectiveness of \systemname{} as a foundation model. Based on \systemname{}, we believe new downstream applications can be developed to improve accessibility and convenience for individuals with disabilities such as sign language recognition.

%

\begin{acks}
We thank the anonymous shepherd and reviewers for their helpful comments. We thank Yupeng Huang for the preliminary validation experiments and all the volunteers involved in this study. This research is supported by HKUST grant R9899, Ministry of Education Singapore MOE AcRF Tier 2 MOE-T2EP20220-0004, and Hong Kong GRF Grant 15206123. Mo Li is the corresponding author.
\end{acks}



\bibliographystyle{ACM-Reference-Format}
\balance
\bibliography{sample-base}



\newpage
\end{document}